\documentclass[epj]{svjour}

\usepackage{amsmath}
\usepackage{graphicx}
\usepackage{fancyhdr}
\usepackage{url}
\def\makeheadbox{{
 }}
\def\mytitle{SFitter: From LHC data back to the MSSM Lagrangian} 
\def\myauthors{M.~Rauch}  
\def\mytype{Contributed Talk}
\def\mysession{Colliders - SUSY Phenomenology}

\DeclareMathOperator{\sgn}{sgn}
\begin{document}
\title{SFitter: Reconstructing the MSSM Lagrangian\\ from LHC data}
\author{\underline{M.~Rauch}\inst{1}
 \and
 R.~Lafaye\inst{2}
 \and
 T.~Plehn\inst{1}
 \and
 D.~Zerwas\inst{3}
}                     
%
%
\institute{SUPA, School of Physics, University of Edinburgh, Scotland
\and LAPP, Universit\'e Savoie, IN2P3/CNRS, Annecy, France
\and LAL, Universit\'e Paris-Sud, IN2P3/CNRS, Orsay, France}
%
\date{}
\abstract{
Once supersymmetry is found at the LHC, the question arises what are the
fundamental parameters of the Lagrangian. The answer to this question
should thereby not be biased by assumptions on high-scale models.
SFitter is a tool designed for this task. Taking LHC (and possibly ILC)
data as input it scans the TeV-scale MSSM parameter space using its new
weighted Markov chain technique. Using this scan it
determines a list of best-fitting parameter points.
Additionally a log-likelihood map is calculated, which can be reduced to
lower-dimensional Frequentist's profile likelihoods or Bayesian
probability maps. 
\PACS{
      {02.50.Fz}{Stochastic analysis} \and
      {12.60.Jv}{Supersymmetric models} 
     } 
} 
\maketitle
\section{Introduction}
\label{intro}
The search for a Higgs boson, or to find an alternative to such a
fundamental scalar, are the main goals of the LHC, which will start
operation next year. However, the Higgs-boson mass is quadratically
divergent with the cutoff scale of the theory, and the question about
possible ultraviolet completions of the Standard Model arises. Such a
completion should also provide an answer for the second unsolved
question in high-energy physics, the existence of cold dark matter.

One possible, and particularly attractive, extension of the Standard
Model is supersymmetry. Its parameter space has been constrained by
various sources over the last years. Collider data from LEP and TeVatron
have put stringent bounds on the masses of the superparticles both via
direct and indirect searches, like the mass of the lightest Higgs boson,
and indirect measurements like the anomalous magnetic moment of the muon
have put further restrictions on the parameter space. Also there is no
one-to-one correspondence of observables and parameters. This is
especially true once we take loop-corrections into account.
At the same time as few assumptions as possible should be imposed. They
should instead be inferred from the available data.

The program SFitter~\cite{sfitter} is designed to solve the task of
mapping up to 20-dimensional highly complex parameter spaces onto a
large set of observables of different quality, which can be highly
correlated. In this proceedings we discuss the techniques and options of
SFitter. The experimentally well-studied MSUGRA parameter point
SPS1a~\cite{sps} is thereby used as an example to illustrate the
features and physical results of this parameter point.

\section{Tools and Techniques}
SFitter can use data input from both high-energy colliders and
low-energy constraints. Possible choices include kinematic endpoints and
thresholds which appear in invariant-mass distributions, mass
differences and masses themselves. Furthermore, it is possible to use
branching ratios and cross sections. The latter ones are normally
associated with large errors, so one would rely on them only when there
are no other types of measurements available. But from a technical point
of view large error bars pose no problem to SFitter. It is even possible
to consider data from e.g.\ ATLAS and CMS measurements separately. 

This data is then compared to the theoretical predictions. Starting from
a parameter point the physical spectrum is calculated by a choice of
three spectrum generators, SoftSUSY~\cite{softsusy},
SuSPECT~\cite{suspect} or
ISASUSY~\cite{isasusy}. Next-to-leading-order cross sections for LHC or a
future ILC are calculated by the program Pros\-pino2~\cite{prospino} and
branching ratios can be included via links to Msmlib~\cite{msmlib} and
SUSY-HIT~\cite{sdecay}. Also the dark matter content of the relic density is
readily available by an interface to micrOMEGAs~\cite{micromegas}. The
communication of parameters and results between the different programs
is performed by the SUSY-Les-Houches-Accord~\cite{slha} data format using
the implementation of SLHAio~\cite{slhaio}. 

Comparing the experimental data $d_i$ and the theoretical prediction
$\bar{d}_i$, where the index $i$ runs over the different data, a
likelihood value is assigned to every point in parameter space. Hereby
we follow the RFit scheme of CKMFitter as described in
Ref.~\cite{ckmfitter} and
assume the theoretical errors $\sigma_i^{\text{theo}}$ as box-shaped.
This scheme interprets theoretical errors as a lack of knowledge on a
parameter. As long as the deviation between theory and experiment is
within the theoretical error, this must not have any influence on the total
likelihood. In combination with the experimental error $\sigma_i^\text{exp}$
the total log-likelihood $\log \mathcal{L} = - \frac{\chi^2}2$ is given by
\begin{equation}
\chi^2 = 
  \begin{cases}
    0 & \text{for $|d_i - \bar{d}_i| < \sigma_i^{\text{theo}}$}\\
    \left(\frac{|d_i - \bar{d}_i|-\sigma_i^{\text{theo}}}%
                {\sigma_i^\text{exp}}\right)^2
      & \text{for $|d_i - \bar{d}_i| \ge \sigma_i^{\text{theo}}$}
  \end{cases}
\end{equation}
The experimental error itself is a combination of three different
sources. All three are considered to be gaussian and therefore
are summed quadratically.
The statistical error is assumed to be uncorrelated between
different measurements. The first systematic error originates from the
lepton energy scale. It is taken to be $99\%$ correlated between
different observables. The second one has its source in the jet energy
scale and is taken with a $99\%$ correlation as well.

To reconstruct Lagrangian parameters a scan of the parameter space
is performed. This is not only important to obtain the correct solution,
or possibly a set of solutions. It is also necessary to extract error
estimates on the parameters from the data.
To accomplish this goal a choice of three different methods is
available, which can be freely combined. The first one is using a fixed
grid. Secondly, Minuit~\cite{minuit} is included as a minimum finder, which
employs a steepest-descent hill-climbing algorithm. The third option uses
the technique of Weighted Markov Chains, which are described in more
detail in the remainder of this section. 

Markov chains are defined as a sequence of points which are the result
of a stochastic process. The special, Markov, property which
characterises them is that the conditional probability of each point
only depends on its direct predecessor, but not on any other previous, or
future, points in the chain. In SFitter the
Metropolis-Hastings~\cite{metropolis_hastings}
algorithm is used for choosing the next point, which works in the
following way. In the first part of the algorithm a new point is
suggested based on the current one using a probability distribution
function (PDF). The PDF can be freely chosen as long as it satisfies the
property that the probability from being at point $x$ and proposing $x'$
is the same as being at $x'$ and proposing $x$. It can for example be
chosen flat, then the Markov-Chain algorithm has no dependence between
points at all and reduces to a simple Monte-Carlo fitting. A good
performance was found using a Breit-Wigner-, or Cauchy-, shaped function.
This type of function has more pronounced tails than a gaussian
distribution and provides a better balance to avoid random-walk
behaviour. The second part of Metropolis-Hastings consists of the
acceptance stage. It is decided whether the suggested point is accepted
or rejected based on a potential, which in our case is $\frac1{\chi^2}$.
So if the log-likelihood of the suggested point is larger
than that of the old one, it is always accepted, else, it is
accepted if the ratio of the two log-likelihoods is larger than a random
number $r$ chosen uniformly between $0$ and $1$. In all other cases the old
point is added to the Markov chain another time:
\begin{equation}
x_{\text{new}} = 
\begin{cases}
x_{\text{sugg}} & 
  \text{if $\log \mathcal{L}(x_{\text{sugg}}) > 
            \log \mathcal{L}(x_{\text{old}})$} \\
& \text{or $\log \mathcal{L}(x_{\text{sugg}})/
            \log \mathcal{L}(x_{\text{old}}) > r$,} \\
x_{\text{old}} & \text{else.}
\end{cases} 
\end{equation}
As a rule of thumb, an acceptance rate of $25\%$ is generally considered
optimal. 
The resulting Markov chain has the property that the density of points
is proportional to the potential, i.e.\ to $\frac1{\chi^2}$ which can
then be used to obtain likelihood maps by binning the points.
To do this binning we have developed a new algorithm~\cite{wmc} based on
previous work by Ferrenberg and Swendsen~\cite{FerrenbergSwendsen}. This
one does not just
take every point with a weight of one, which corresponds to the
classic way. As the function we want to plot is the same which has been
used in the acceptance decision, this information can be reused. We
therefore weight the points with the potential function taking care that
no double-counting is done and points where the likelihood vanishes are
properly taken into account.

It was shown~\cite{baltzgondolo} that Markov-chain techniques can scan
high-dimensional parameter spaces very efficiently. They have been
extensively applied to constrain the MSUGRA parameter space from current
experimental precision data~\cite{ben,leszek}.

\section{Reconstructing SPS1a}

With these possibilities SFitter provides the relevant frequentist or
Bayesian results in three steps: First, a fully exclusive log-likelihood
map of the complete parameter space is calculated. In a second step the
best local likelihood maxima are obtained from the map and ranked
according to their likelihood. Last, the map is projected onto
lower-dimensional spaces down to one-dimensional distributions. This can
be done in both a Frequentist way, which yields a profile likelihood,
and in a Bayesian way. Here the unwanted dimensions are marginalised
away and a prior needs to be specified for this.

In general, no specific model for supersymmetry breaking should be
assumed a priori. Instead this should be inferred from the data.
However, to demonstrate the features of SFitter we consider here the MSUGRA
scenario using the parameter point SPS1a. Besides the MSUGRA parameters
$m_0$, $m_{1/2}$, $\tan\beta$, $A_0$ and $\sgn\mu$ also the top-quark
mass $m_t$ is taken as a free parameter because of the large uncertainty
on its value. A smeared dataset is generated from the measurements which
can be performed in this parameter point. Using 30 individual chains with
20000 points each a Weighted Markov run is used to produce the
likelihood map. For the best points an additional Minuit fit is
performed to find the exact position of the minimum. With this procedure
we find four distinct minima as shown in Table~\ref{msugra:table}. The
best point corresponds indeed to the true solution where the deviations
turn out to be compatible with the error bars. The second solution is
given by very similar values for the continuous parameters and a flipped
sign of $\mu$. Solution three shows a distinct maximum, and the last
one again differs from the previous one only by $\sgn\mu$. For these
last two solutions the trilinear
parameter takes a large positive value together with a slight shift in
the top-quark mass. If the latter was kept fixed this distinct maximum
would be much less pronounced. In the last line of 
Table~\ref{msugra:table} the corresponding errors on the parameters are
printed.
\begin{table}
\begin{tabular}{l|rrrrrr}
     $\chi^2$&$m_0$ &$m_{1/2}$ &$\tan\beta$&$A_0$&$\mu$&$m_t$ \\ \hline
     0.09  &102.0 & 254.0 & 11.5 & -95.2  & $+$ & 172.4 \\
     1.50  &104.8 & 242.1 & 12.9 &-174.4  & $-$ & 172.3 \\
     73.2  &108.1 & 266.4 & 14.6 & 742.4  & $+$ & 173.7 \\
    139.5  &112.1 & 261.0 & 18.0 & 632.6  & $-$ & 173.0 \\
          \dots \\\hline
    errors & 2.17 &  2.64 & 2.45 &  49.6  &     &  0.97
     \end{tabular}
\caption{SFitter output for the SPS1a point in MSUGRA. List of the
best-fitting parameter points with associated log-likelihood. The last
line denotes the corresponding errors for the best-fitting
point. All masses are in units of GeV.}
\label{msugra:table}
\end{table}

\begin{figure}
\includegraphics[width=0.4\textwidth]{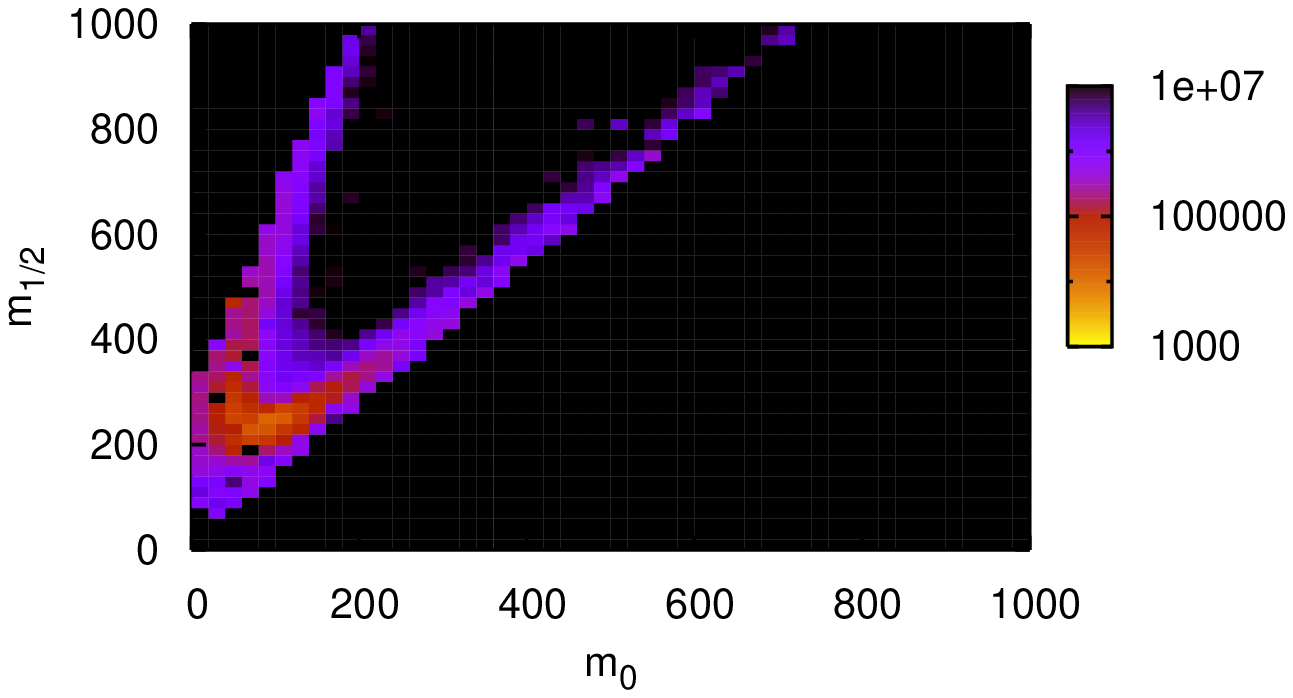}\\
\includegraphics[width=0.4\textwidth]{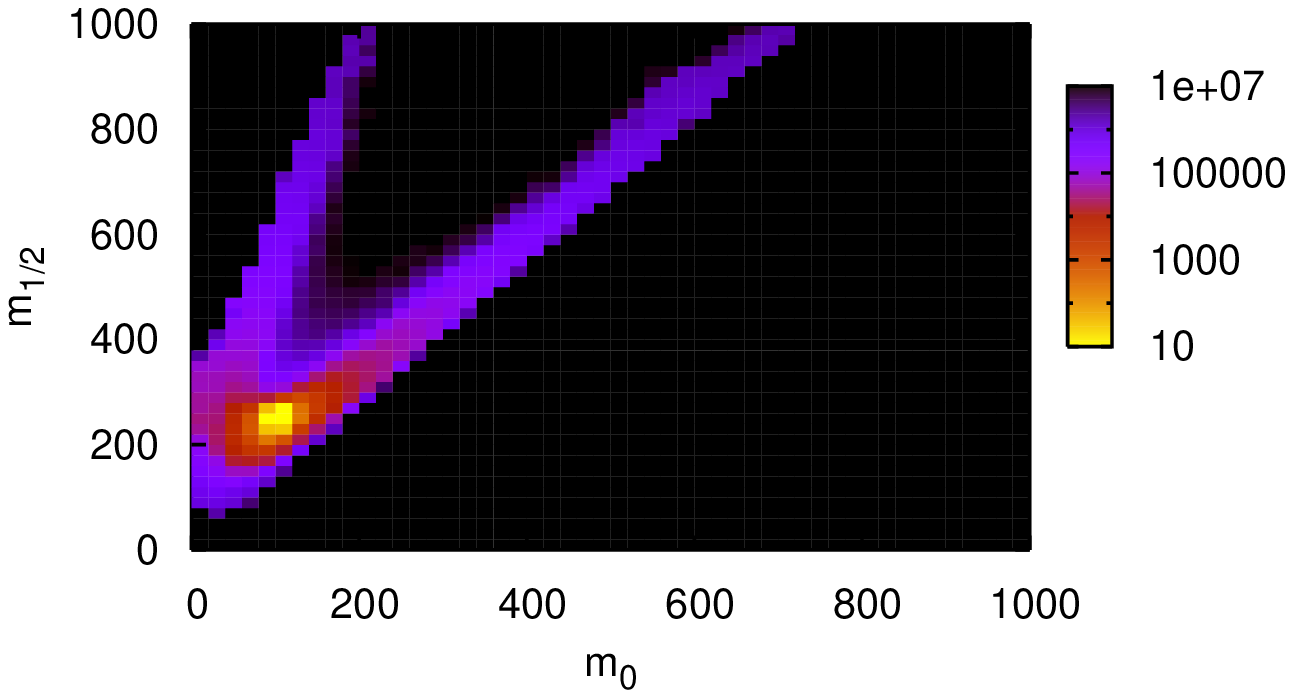}
\caption{SFitter output for the SPS1a point in MSUGRA. 
Two-dimensional likelihood maps of the $m_0$-$m_{1/2}$ plane. Upper:
Bayesian marginalised plots. Lower: Profile likelihoods. 
All masses are given in GeV.}
\label{msugra:plots_2d}
\end{figure}
Fig.~\ref{msugra:plots_2d} shows the corresponding plots of the likelihood
map projected onto the $m_0$-$m_{1/2}$ plane. The peak is clearly at the
right position in both the marginalised plot and the profile likelihood.
The resolution of the plots is too coarse so the different maxima are
merged in a single bin and cannot be resolved here.  From there it
extends in two branches, which reflect the fact that extracting masses
from kinematic endpoints involves quadratic equations. 

Strictly speaking, the usual set of MSUGRA parameter is not purely
high-scale, as it contains the weak-scale quantity $\tan\beta$ which
explicitly assumes radiative electroweak symmetry breaking. This
can be replaced by the mass parameters $B$ and $\mu$~\cite{ben}
which appear as $B\mu$ in front of mixed terms of the type $H_1^0
H_2^0$. $\mu$ can then be eliminated by the requirement that the correct
low-energy $Z$-boson mass is reproduced.
This distinction is important for the marginalised plots, where a prior 
as a measure in the parameter space must be specified.
\begin{figure}
\includegraphics[width=0.4\textwidth]{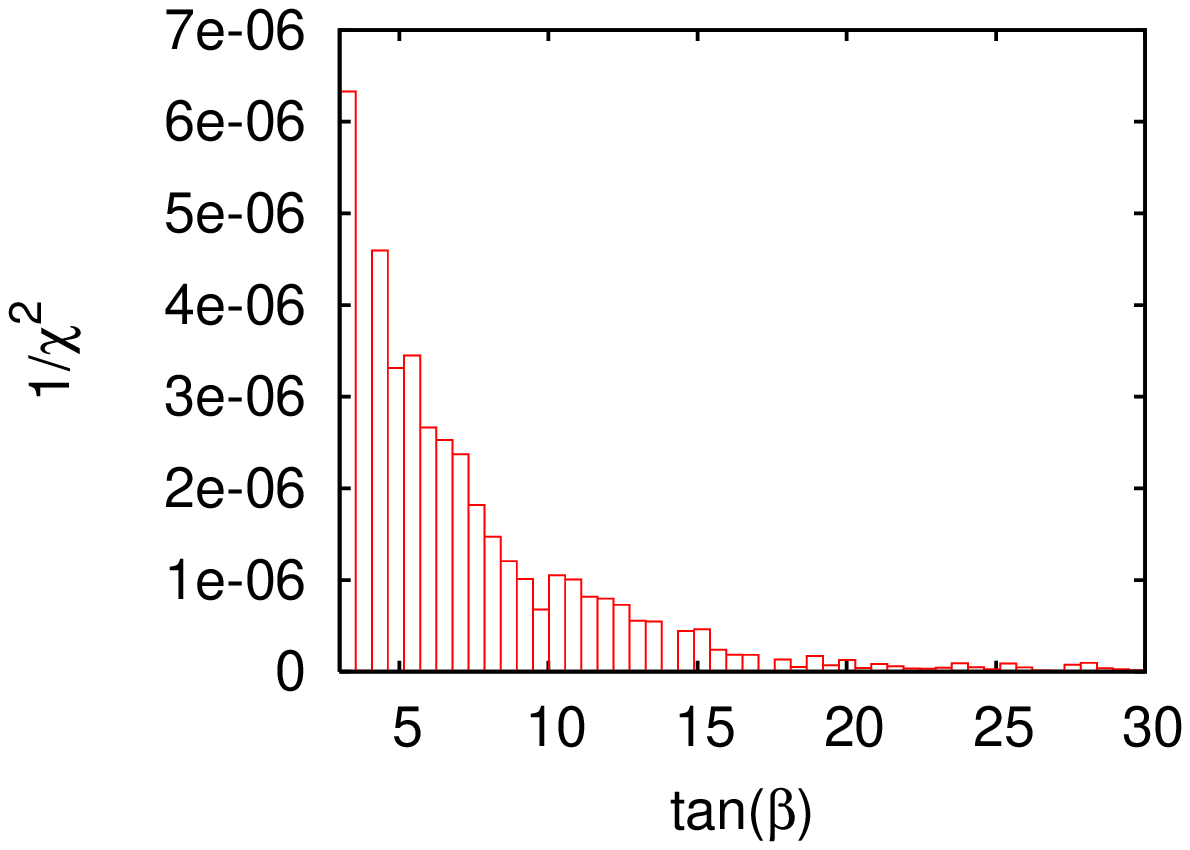}\\
\includegraphics[width=0.4\textwidth]{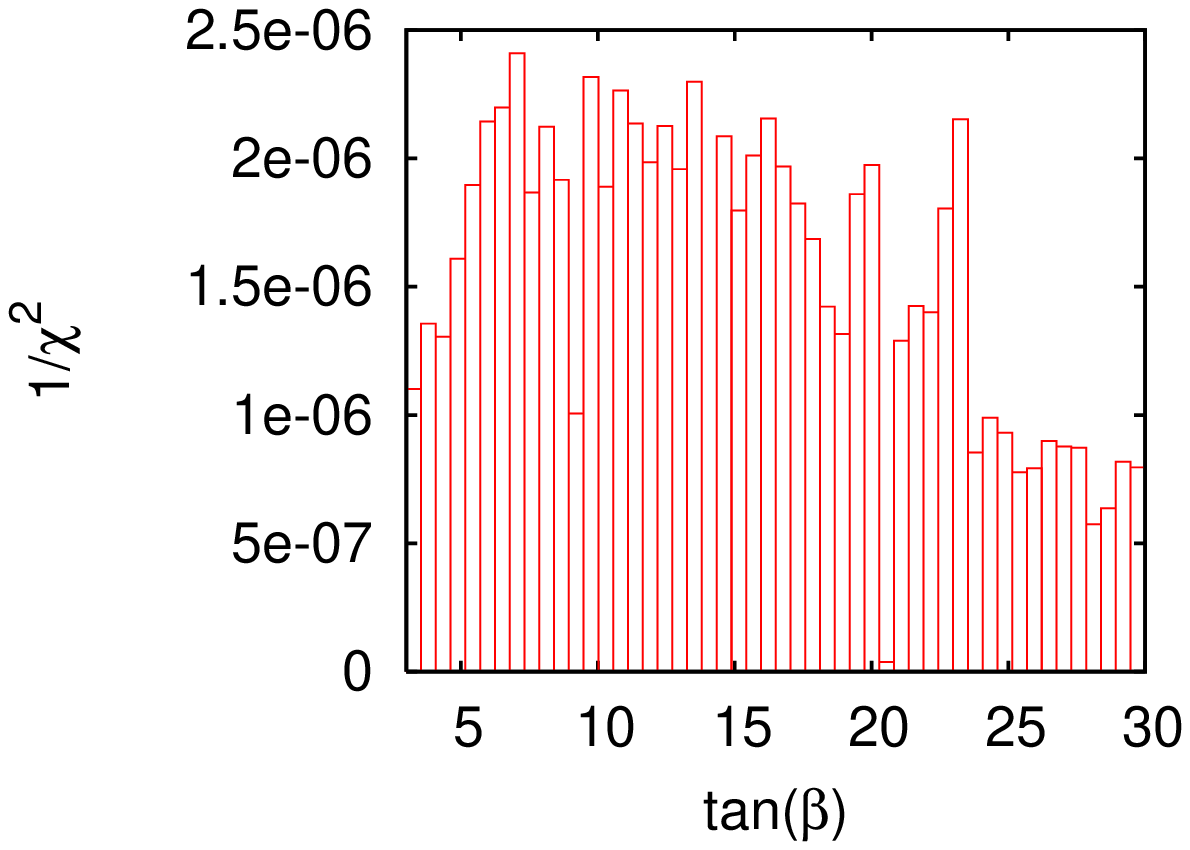}
\caption{SFitter output for the SPS1a point in MSUGRA. 
One-dimensional Bayesian marginalised plots for $\tan\beta$. 
Upper: High-scale prior flat in $B$.
Lower: Prior flat in $\tan\beta$.}
\label{msugra:plots_tb}
\end{figure}
Fig.~\ref{msugra:plots_tb} shows the different results of the two
choices. In the upper plot a purely high-scale model is chosen as has
been done in the previous plots. The prior is taken to be flat in $B$.
This corresponds to a prior in $\tan\beta$, which falls of as
$1 / \tan\beta^2$ in leading order. This behaviour is clearly visible in
the plot. It is dominated by noise and the prior and as small values of
$\tan\beta$ as possible are preferred. In the lower plot a prior which
is flat in $\tan\beta$ has been chosen. Here the plot still shows a
significant noise, but the maximum is in the correct place. 
For profile-likelihood plots this choice does not making a difference in
the resulting likelihood. It does however have an indirect influence via
the Markov chain scanning. For the PDF a measure in the parameter space
must be defined which influences the probability for suggesting a point.
A bad choice of the PDF can then lead to a bad coverage of the parameter
space.

\section{Dark Matter}

A very important clue for Physics beyond the Standard Model is the
existence of cold dark matter, which in the SM cannot be created in the
right quantity. Supersymmetry provides an ideal candidate to solve this
problem. The lightest neutralino as the lightest supersymmetric particle
is stable, massive and only weakly interacting. That it is indeed
responsible for the dark matter content of the universe is a hypothesis
which should not just be simply assumed and used in the fits. Using
future collider data it is possible to test it. After the Lagrangian
parameters and their associated errors have been extracted, the relic
density can then be computed and compared to the experimental data. 
In this short analysis we ignore the fact that SPS1a produces a too
large density which is excluded and would overclose the universe, since
we are not interested in the central values but in the principal
dependence of the parameters and the propagation of
the errors. 

Using the information on the parameters obtained in the previous section
we use micrOMEGAs to calculate an estimate of the relic density
$\Omega_\text{CDM, SPS1a} h^2 = 0.1906 \pm 0.0033$. The amount decreases
with smaller $m_0$ and larger $\tan\beta$, while the dependence on
$m_{1/2}$ and $A_0$ is rather weak. This can then be compared to the
experimental value, which is derived from the measurement of the
fluctuations of the cosmic microwave background by WMAP~\cite{wmap}. The
current best value is $\Omega_\text{CDM, exp} h^2 = 0.1277 \pm 0.008$.  
Adding possible future ILC measurements would mean an additional
error reduction in the extracted value by a factor of 10. So the
accuracy which can be derived from collider data alone is well
compatible with the experimental precision and testing the hypothesis
that the lightest neutralino is responsible for the dark matter content
of the universe viable.

\section{Conclusions}

If new physics is discovered at the LHC, the crucial task will be to map
the possibly strongly correlated observables onto a high-dimensional
weak-scale Lagrangian. SFitter with its new weighted Markov chain
technique has been designed to accomplish such a task. It produces both
a list of best points and a fully-dimensional exclusive likelihood map
as output. The dimensionality of this map can then be reduced using
either Frequentist or Bayesian techniques, which yield a profile
likelihood or marginalised plots, respectively. The latter depend on the
choice of priors, which can have a significant effect when the data
cannot constrain parameters sufficiently and demonstrate Bayes\-ian volume
effects. After the reconstruction the parameter set can be used for a
prediction of the relic density, which can then be compared to the
experimental value. We find that for SPS1a both values have comparable
errors. SFitter is despite its name not limited to supersymmetry. It can
and will be used to study further problems involving high-dimensional
parameter spaces.


\end{document}